\documentclass [twoside]{article}
\usepackage{epsfig}
\author{ P.~Kulinich \\
\small MIT, Cambridge, MA, USA;  \ \ kulinich@mit.edu \\
S.~Basilev \\
\small VBLHE, JINR, Dubna, Russia \\
V.~Krylov \\
\small LIT, JINR, Dubna, Russia
}

\title{A novel String Banana Template Method for Tracks Reconstruction in High 
Multiplicity Events with significant Multiple Scattering and its Firmware Implementation
}

\begin{document}
\markboth{SBTM Firmware}{SBTM Firmware}

\maketitle

\begin{abstract}
Novel String Banana Template Method (SBTM) for track reconstruction in difficult conditions
  is proposed and implemented for off-line analysis of relativistic heavy ion collision events.
 The main idea of the method is in use of features of ensembles of tracks selected by 3-fold 
coincidence. Two steps model of track is used: the first  one -- averaged over selected 
ensemble and the second -- per event dependent. It takes into account Multiple Scattering (MS)
 for this particular track. 

SBTM relies on use of stored templates generated by precise Monte Carlo simulation, 
so it's more time efficient for the case of 2D spectrometer. All data required for track 
reconstruction in such difficult conditions could be prepared in convenient format for fast use. 
Its template based nature and the fact that  the SBTM track model is actually very close to the
 hits implies that it can be implemented in a firmware  processor. In this report a block 
diagram of firmware based pre-processor for track reconstruction in CMS-like Si 
tracker is proposed.

\end{abstract}

\section{Introduction}
Some modern experiments have to deal with tracks in  high multiplicity events in spectrometers
 with finite granularity detector and often in a non-uniform magnetic field and in the presence
 of significant MS. 
Considerable MS up to ten mrad per silicon detector plane, energy loss at low momenta and issues
 like clusterization and large incident angles potentially create difficulties in pattern 
recognition and track fitting in high multiplicity events even for primary tracks. 
\subsection{Effect of Multiple Scattering}

Usually for pattern recognition local methods are used and fitting of track's parameters in 
the case of non negligible MS is accomplished by a modification of linear 
Least Square Method (LSM).
 In any case Covariance matrices of different sizes should be prepared and inverted during 
fitting procedure.   Iterative procedures are required in the non-linear cases. Long 
combinatorial loops inevitable for local pattern recognition methods require much 
computation time at the fitting stage. 

In the SBTM there is another approach to handle effect of MS. There are no matrices and 
fitting procedure is quite different from standard -- the best track parameters are achieved 
for the geometrical model most close to trajectory hits. There are few more parameters for 
SBTM track model and they depend on particular set of multiple scatterings for a given track.
 They provide additional (per event) corrections for track parameters. 

Capabilities of the SBTM are demonstrated by comparison  of track parameters resolutions
 to results received by LSM method for toy model spectrometer. Proposed global method has 
good pattern recognition features for primary tracks in high multiplicity events because 
of narrow search windows and a priori known momentum. Some basic characteristics of 
implemented into C++ SBTM track reconstruction software for environment of heavy ion 
collisions are provided.

\section{Features of SBTM}

The main idea of the method is in use of ensembles of particle trajectories with 
three fixed points for templates accumulation \cite{Kulinich_}. 
Two steps model of track is used: the first  one 
-- averaged over ensemble (although different from ``standard'') and the second -- 
per event dependent (takes into account MS for this particular track).
  
\begin{figure}  [h]
\begin{center}
\mbox{\epsfig{bbllx=0mm,bblly=0mm,bburx=194mm,bbury=270mm,%
height=18cm,width=14cm,file=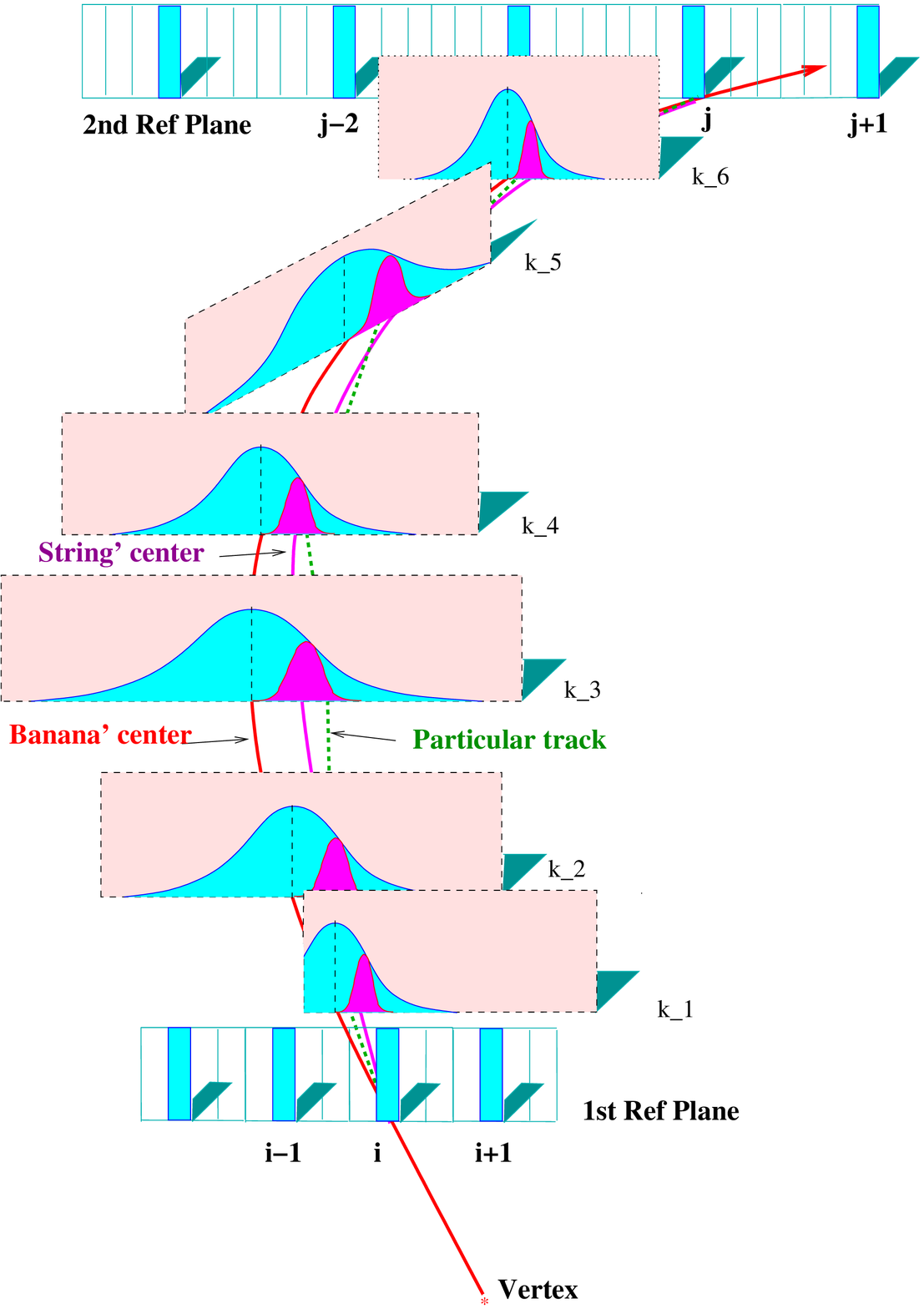,clip=}%
}
\end{center}
\vspace*{8pt}
\caption{ 
{\small $Banana$ $(i,j)$ ``bell road''  and a particular $String$ 
-- more narrow ``bells'' inside, for ensemble with ``narrow'' momentum 
distribution which  originated from $V_{0}$ and
pass  through the same pixels ``$i$'' in FRP and ``$j$'' in SRP (``3-fold coincidence'').
}}
\label{SBTM_log}
\end{figure}

Ensembles of tracks from given vertex point which pass through particular 
pixels (strips) \{i,j\} in two Reference Planes (RPs) are formed. For such ensemble 
(which geometrical image  has a $Banana$-like shape in magnetic field, 
see Fig. ~\ref{SBTM_log})
 all necessary values corresponding to each internal planes (as centre of $Banana$s, 
their widths, angles, lengths an so on)  could be  saved. 

At first track recognition stage one check different combinations of \{i,j\} signals in RPs 
and select such which has proper number of signals in all (or in almost all planes) inside 
of $Banana$ window. At the second stage relative positions of signals inside $Banana$ are 
checked and if they are inside  a more narrow window ($String$) track candidate is recognized. 
Pattern recognition is accomplished in two relatively simple stages and there is no traditional 
fitting procedure for obtaining of track parameters (no hard computations). Because of such 
conditional correlation on MS  the actual width of template's search window ($Banana$) 
on the 1-st stage
 of pattern recognition becomes relatively small. On the second stage the 
search window ($String$) is ``shrunk'' again few times (about 4...6).
 This simplifies and accelerates (the global in nature)
 pattern recognition process, by avoiding large combinatorial trees search. 

SBTM relies on use of stored templates, generated by precise Monte Carlo simulation prepared 
in advance. All necessary data required for track reconstruction could be prepared for fast use.
 SBTM track model is actually very close to trajectory hits (even for low momentum tracks), 
so it can be implemented in a firmware  processor. For high multiplicity events without 
external track seeds number of combinations \{i,j\} to be checked becomes large, so hardware
  or firmware processor which could accomplish this task often in parallel and much faster 
could be used as a trigger processor or pre-processor for subsequent off-line analysis on 
standard computers (providing selected tracks candidates for final qualitative checking).

\subsection{Comparison to LSM results}
For the purpose of demonstration, the SBTM method was applied to a two dimensional toy model
 spectrometer as used in \cite{Lutz_}. It consists of four high resolution silicon detectors uniformly 
distributed over 12 cm track length followed by thirteen gas detectors again at equal spacing
 over 120 cm track length. All planes are parallel. Properties of this setup are summarized 
in Table 1 of \cite{Lutz_}. The spectrometer is placed in a 1T transverse magnetic field. The 
thickness of Si planes is 0.4 \% of radiation length, for gas detectors it's 0.1 \%. 
The spatial resolutions are 5 and 200 $\mu m$ respectively.

\begin{figure} [ht]
\begin{center}
\mbox{\epsfig{file=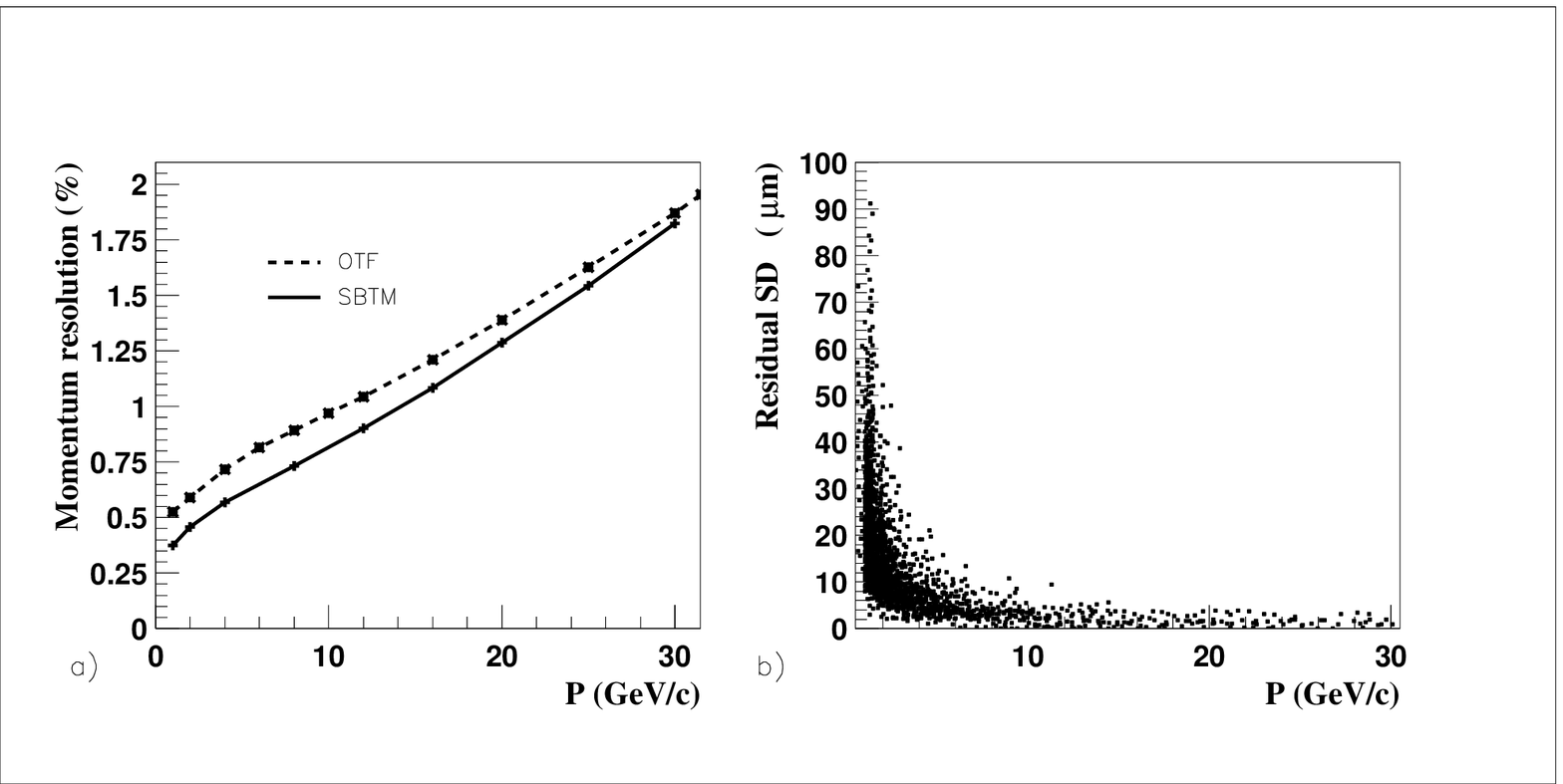,height=8cm,width=14.5cm,%
bbllx=5mm,bblly=5mm,bburx=190mm,bbury=85mm,clip=}%
}   
\end{center}
\vspace*{8pt}
\caption{{\small $a)$ Momentum Resolution for two methods described in text.
 $b)$  Space precision of the SBTM model --
 residual standard deviation (for ``ideal'' space resolution case). 
It  shows how close  SBTM model is to actual hits.}}
\label{Imp_par}
\end{figure}

The optimal track fitting (OTF) used in \cite{Lutz_} reproduces the results 
of the global fit \cite{Wolin_}.
 On Fig.~\ref{Imp_par}a) momentum resolutions for SBTM and OTF methods are shown. 
SBTM could be used for vertex finding also. Pattern recognition including vertex 
finding was done initially for  few vertex points along Y axis to find the proper vertex.
 To achieve good resolution, fine track tuning-rotation of template's ends inside strip 
width  for FRP (layer 4) and inside of $3\cdot \sigma$  for SRP (last layer) was  performed.
 In that case two additional parameters were used for two $String$s (one $String$ before
 FRP and the second - between FRP and SRP). 
SBTM potentially provides better track parameter's resolutions than method equivalent 
to LSM. Fig.~\ref{Imp_par}b) shows how close track model is from actual hits;
 it is received for the 
ideal case when each plane has very good space resolution (less than few $\mu$m).

\subsection{Implementation for PHOBOS spectrometer }

\begin{figure} [ht]
\begin{center}
\mbox{\epsfig{file=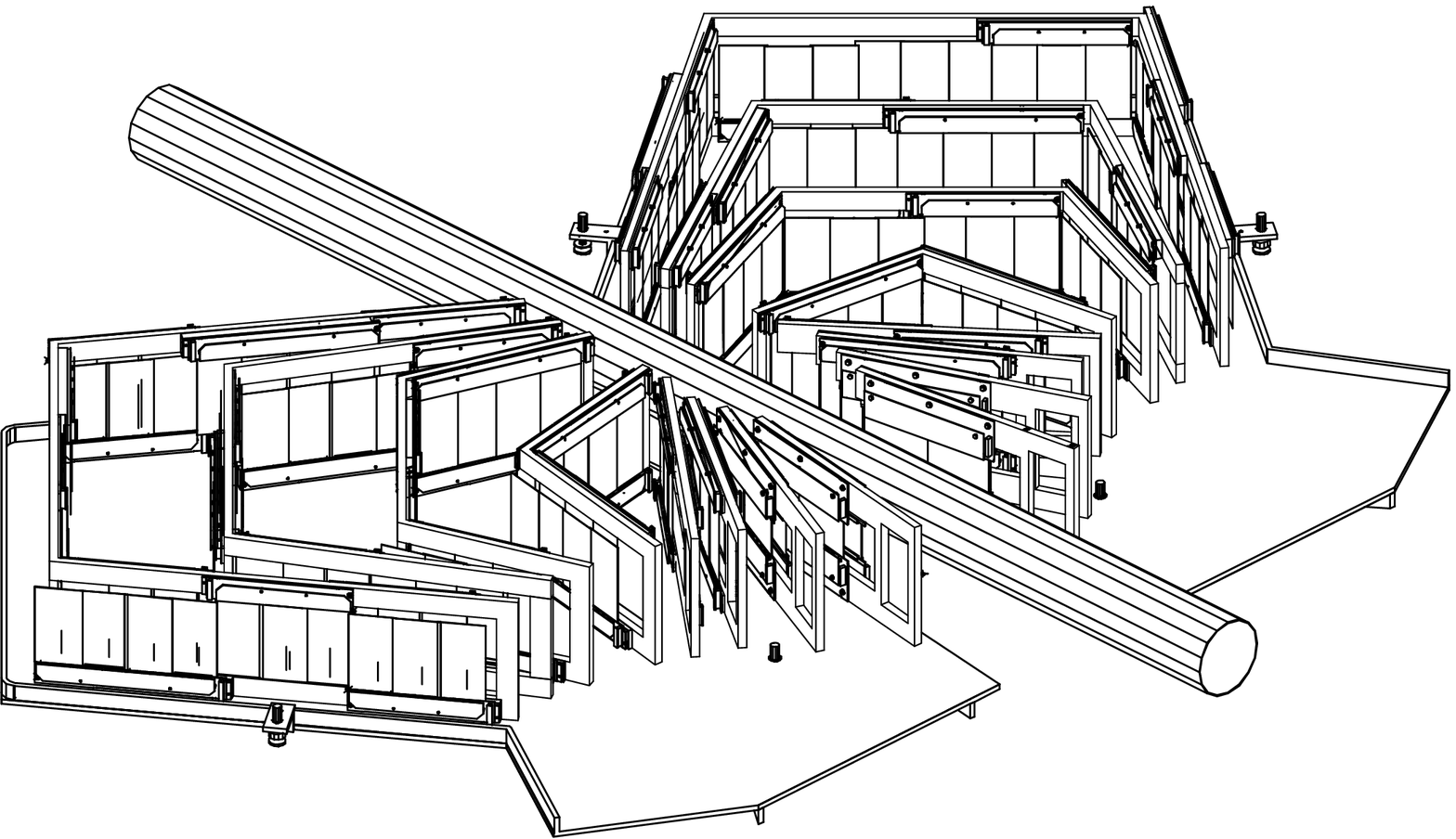,height=8cm,width=14.7cm,%
bbllx=0mm,bblly=0mm,bburx=222mm,bbury=122mm,clip=}%
}   
\end{center}
\vspace*{8pt}
\caption{{\small
Si based PHOBOS spectrometer. The dimensions of one arm $\simeq$(8x80x80) $cm^{3}$. Total number of 
analog read-out channels is about 108k.}
}
\label{PH_set}
\end{figure}

The PHOBOS  Spectrometer \cite{PHOBOS_} (for relativistic heavy ion experiment at RHIC) has two 
arms (see Fig.~\ref{PH_set}) with Si detectors, located on the opposite sides of the beam pipe. 
The first few nearest to the pipe planes of each Spectrometer arm reside in low magnetic
 field region, while the subsequent planes are placed in a  vertical field with maximum of 2~T.
  The dimensions of the silicon sensor pads range from 0.4~mm to 1.0~mm in the bend direction. 
Additional segmentation in the vertical direction  is provided: first four layers (closest to 
the interaction region) have 1 mm height pads; for the  rest layers heights progressively  
increase  from 6 to 19~mm. Initial objective of tracking was low transverse momentum charged
 particles.

Templates are generated  for different vertexes with step 5~mm along Z axis (along  beam pipe).
 Ensembles of tracks with common vertex  which pass through  a particular pair of pixels 
\{i,j\} in two Reference Planes (RPs) are used for calculation of mean values and 
some $\sigma$s 
for different templates parameters. In our case the last plane before magnet and one of 
the last plane in magnet are used as RPs.

\begin{figure} [ht]
\begin{center}
\mbox{\epsfig{file=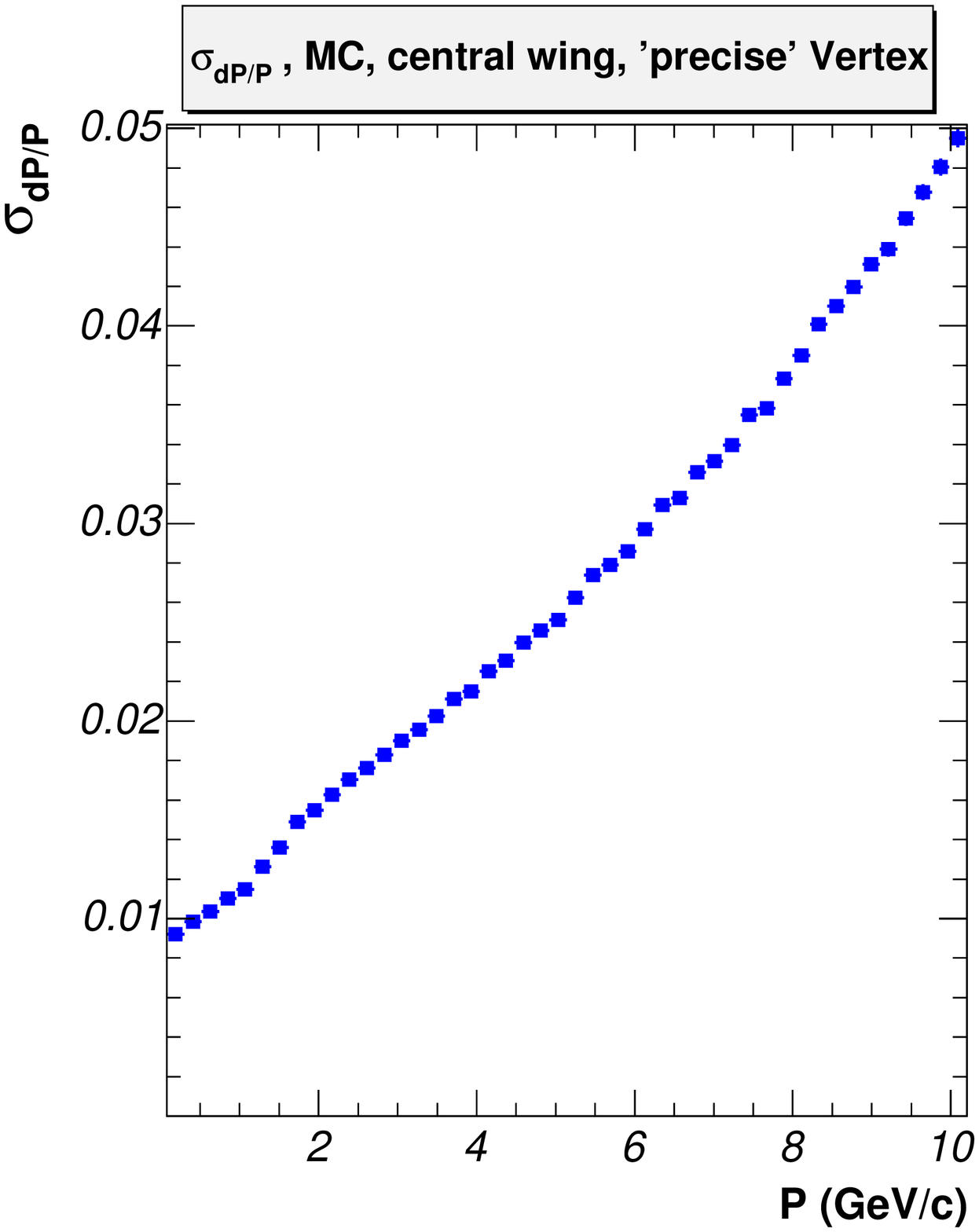,height=5.5cm,width=9cm,%
bbllx=5mm,bblly=12mm,bburx=190mm,bbury=230mm,clip=}%
}   
\end{center}
\vspace*{8pt}
\caption{{\small
Momentum Resolution for $\pi$ in the central wing with exactly known vertex position.
}}
\label{PH_prez}
\end{figure}

SBTM has high efficiency in the high multiplicity Monte Carlo events ($dN/d\eta\sim5000$) 
for track reconstruction in the wide range of momenta, starting from the lowest possible
 values for tracks which cross all planes ($\sim$80~MeV/c for $\pi$,
 see Fig.~\ref{PH_prez}).

\begin{figure} [ht]
\begin{center}
\mbox{\epsfig{file=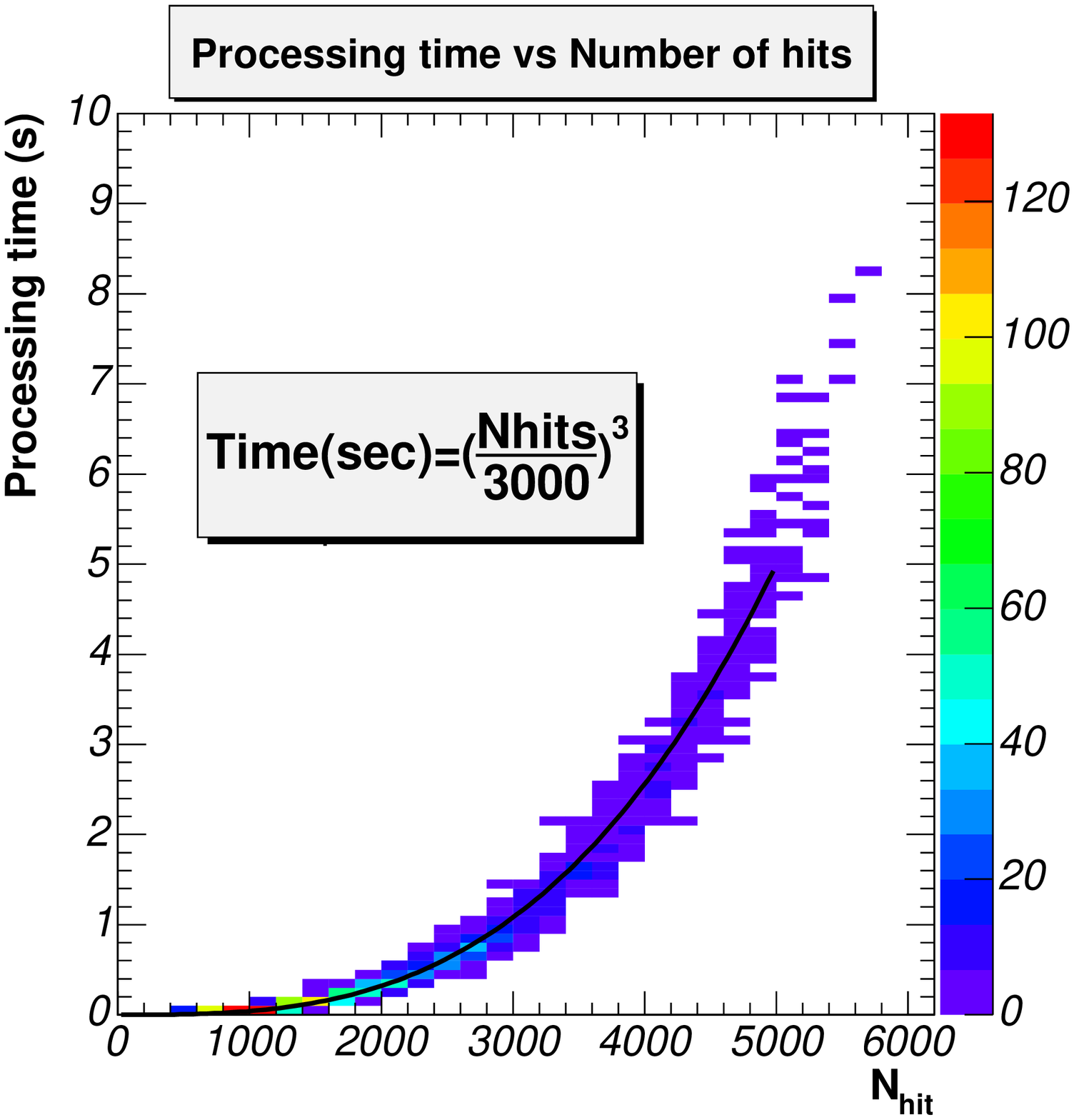,height=8cm,width=5.5cm,%
bbllx=0mm,bblly=20mm,bburx=200mm,bbury=190mm,angle=270,clip=}%
}   
\end{center}
\vspace*{8pt}
\caption{{\small
Processing time on Xeon 3~GHz as a function of number of hits (two arms). 
}}
\label{PH_tim}
\end{figure}

This pattern recognition method and use of dE/dx information for each channel permit to 
reconstruct close tracks which share up to $3/4$ of their hits. For real data occupancies 
reach the level of 25 \% in high $\eta$ regions of spectrometer. 
On Fig.~\ref{PH_tim} processing time (which includes pattern recognition and ghost cleaning) 
is shown as a function of number of the hits in the arms for real data in 
(Au + Au) collisions at  $\sqrt{s_{NN}}=200$~GeV.

\section{Firmware Realization}

\begin{figure} [ht]
\begin{center}
\mbox{\epsfig{file=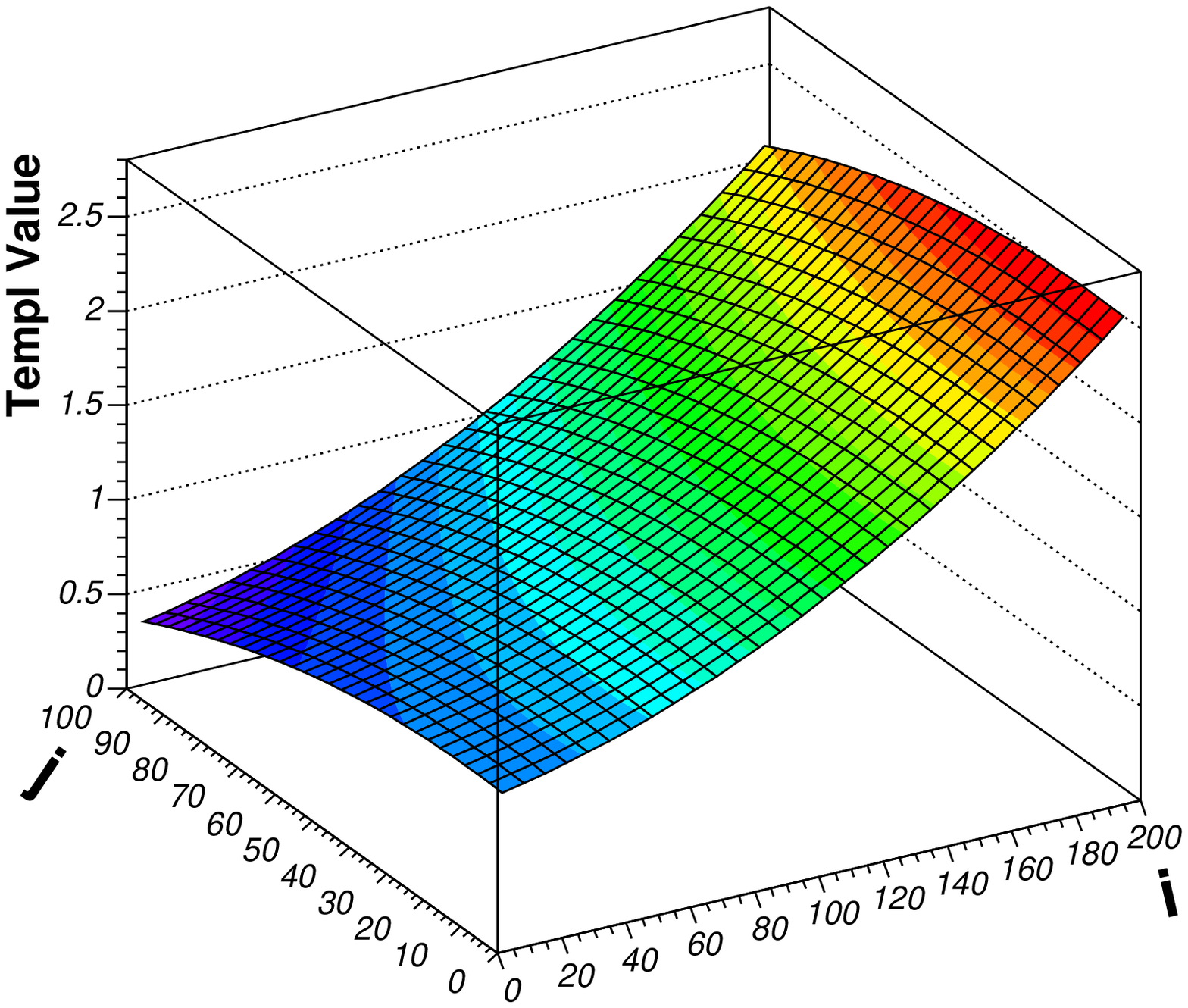,height=5cm,width=8.5cm,%
bbllx=12mm,bblly=10mm,bburx=177mm,bbury=148mm,clip=}%
}   
\end{center}
\vspace*{8pt}
\caption{{\small
Example of smooth  Template surface and the grid on it. Inside  the  grid cell 
template parameters could  be  well approximated by  linear interpolation. 
}}
\label{templ_s}
\end{figure}

Templates volume could be large if data are saved for each possible combination of \{i,j\}
 reference channels. But due to the fact that these 3D surfaces are quite smooth 
(as on Fig.~\ref{templ_s}) they could be parametrized as 2D functions (or slices of 2D histogram) 
and in that case the size of templates could be relatively small.
 Another option is to save template values for the grid points as 
on Fig.~\ref{templ_s} and then use linear interpolation for internal points. 
In that case the fast look-up tables with required address space become affordable.

\begin{figure}  [h]
\begin{center}
\mbox{\epsfig{bbllx=0mm,bblly=0mm,bburx=190mm,bbury=255mm,%
height=18cm,width=14.cm,file=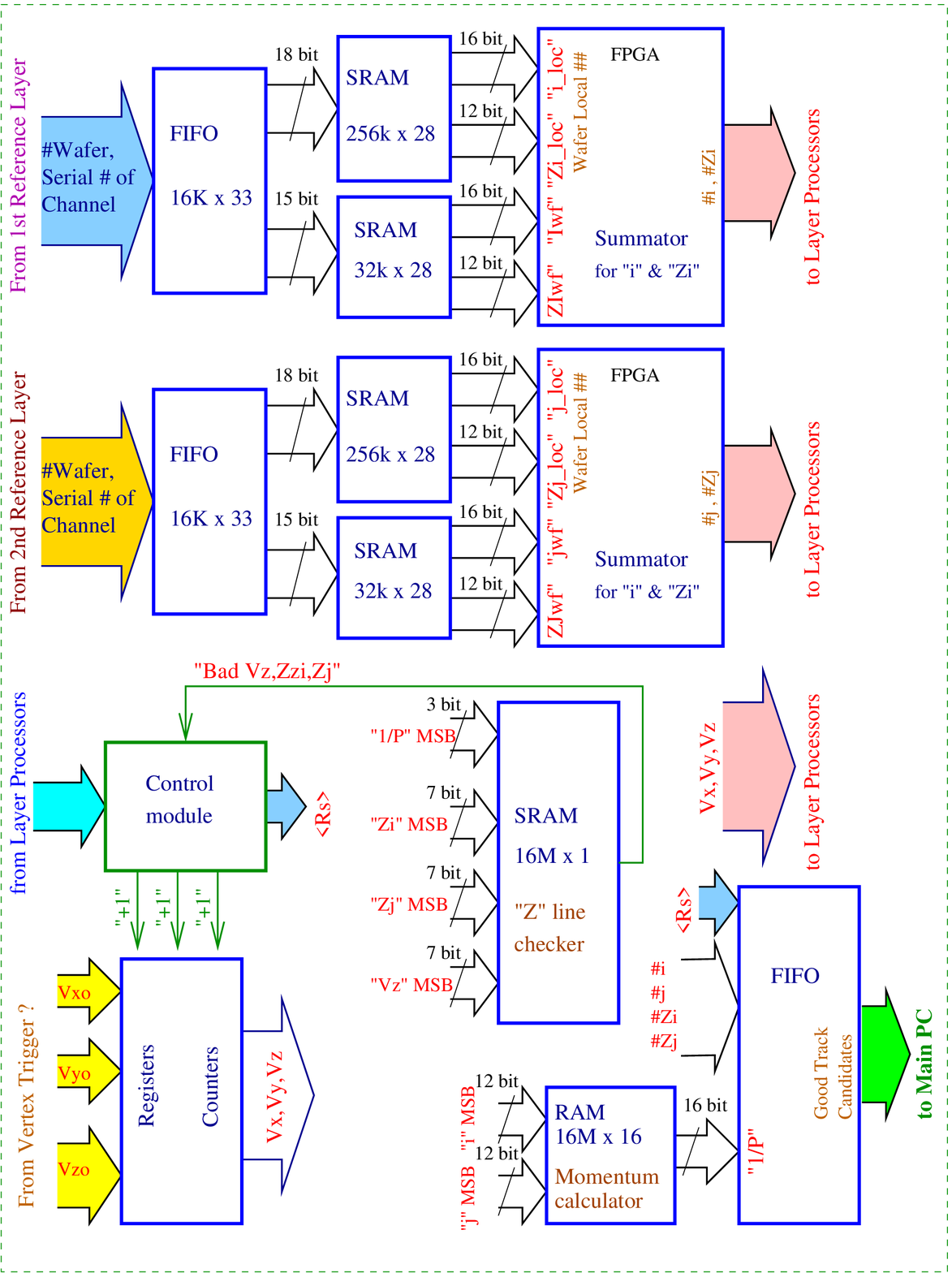,clip=}%
}
\end{center}
\vspace*{8pt}
\caption{{\small 
Block diagram of Master module.
}}
\label{SBTM_Master}
\end{figure}

\begin{figure}  [h]
\begin{center}
\mbox{\epsfig{bbllx=0mm,bblly=0mm,bburx=200mm,bbury=255mm,%
height=18cm,width=14.cm,file=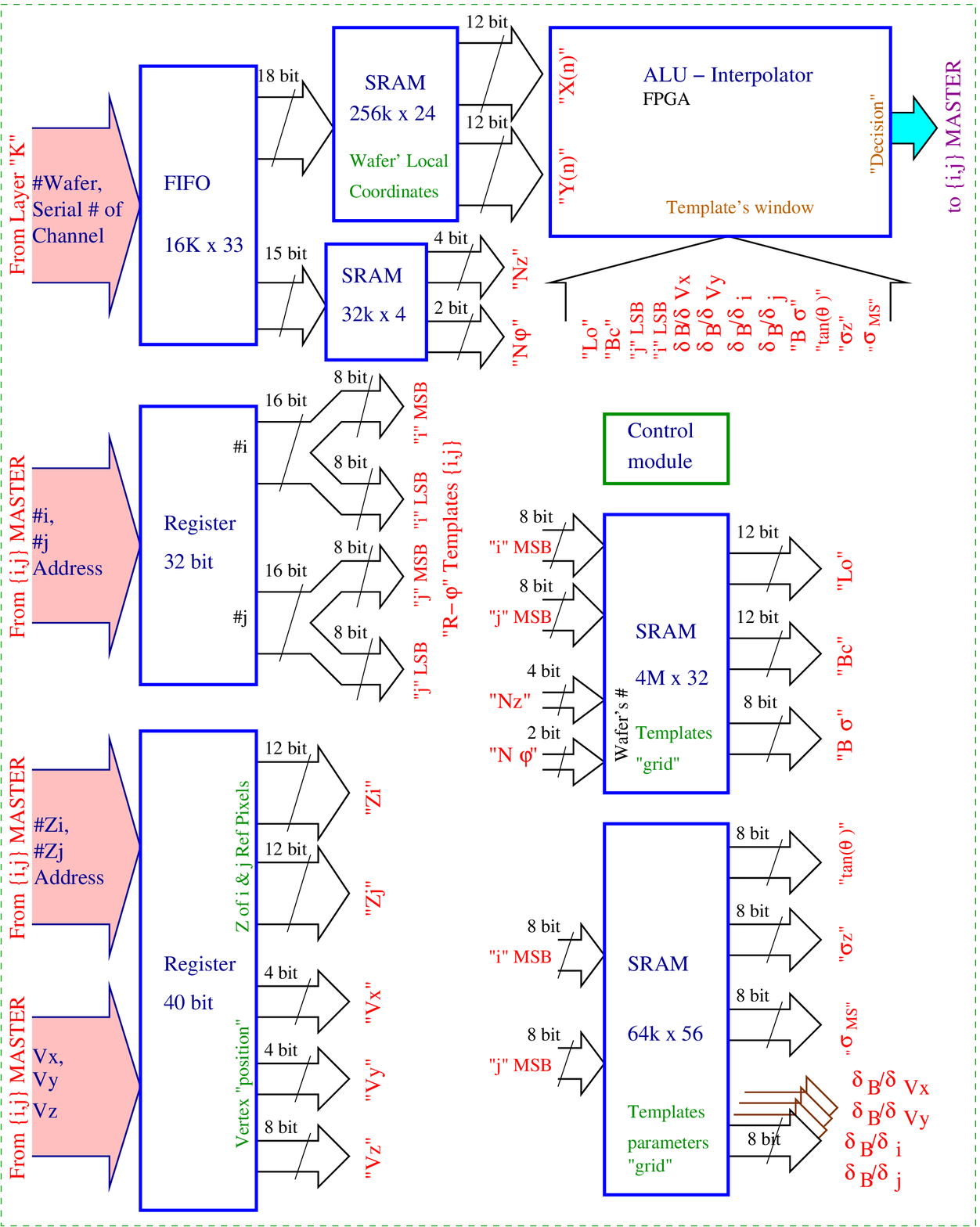,clip=}%
}
\end{center}
\vspace*{8pt}
\caption{{\small 
Block diagram of BPM module.
}}
\label{SBTM_BMP}
\end{figure}

 In this paper a block diagrams of firmware based pre-processor which uses grid 
templates with relatively fast interpolation (using 8 bit multiplication look-up 
tables)  is presented and basic features of implementation are described. 
It consists of a master module BMM (Fig.~\ref{SBTM_Master})
 and of number of processing modules 
BPMs (one per layer or part of the layer; Fig.~\ref{SBTM_BMP}) 
 connected  by fast custom bus.
During initialization process BPMs receive corresponding plane's templates data 
in convenient format.  All local coordinates and widths are converted into fixed 
float format with small bins. 

At reconstruction stage the input data from spectrometer events  are transmitted
 in parallel in raw format by dedicated computer nodes. In the processor's modules
 this information is converted into reference channel numbers \{i,j\} 
(BMM) and fixed float format (BPMs). Detector planes are subdivided into small 
sub-frames (as Si wafers for example) which have their relative numbering. 
Event data buffers for processing are saved in FIFOs. If there is fast 
Vertex processor then its information could be received by BMM, otherwise 
Tracking Processor should check different vertex positions to find one with 
optimum number of tracks and their deflections from the $String$s.

Master module receives information from two reference planes. It loops over all hits in 
both RPs for all possible combinations of \{i,j\} in the mode  without external designation. 
BMM has a look-up memory for checking that three points: vertex and two reference pixels 
\{i,j\} in two RPs are close to the strait line in the Z plane (along magnet field).
 It could decrease the number of combinatorial loops in BPMs by factor $\sim300$ 
(for geometry similar to CMS Si tracker). 

After preliminary Z selection of the pairs of hits  BMM provides \{i,j\} indexes
 for  template's grid addresses for processor modules. It asks BMPs to verify the 
existence of signals inside the $Banana$ windows with indexes \{i,j\}. If there are
 enough signals in all planes (conformation signals from all BPMs, or almost all 
for realistic detection efficiencies) inside that particular $Banana$, then this 
\{i,j\} could be a good candidate in the case of low occupancy event.

  At this stage it's not yet final track candidate. One need to check that signals
  in all planes inside such $Banana$ could be localized in much more narrow window 
-- $String$ with relative deflection $Rs$ ($String$ parameter). Such decision is based 
on  received information about relative deflection of hits from $Banana$'s centres
  in each plane (from  all  layer  processors; in 1D histogram' position code).

BMM  accumulates  the histogram of $Rsk$ and if there is such combination
 of signals which gives histogram maximum over some threshold --
 a new \{i,j\} track candidate'  information and additional $String$ parameter
 $<Rs>$ are saved into output buffer. At this stage the momentum value of candidate
 taken from templates is corrected based  on $<Rs>$. Such $String$ search at the
  the second stage of patten recognition is accomplished by pipelined 
firmware architecture of BMM.

For the case of events with 1000 tracks  in acceptance of such pre-processor one 
could expect processing time to be about (50...100) ms if the vertex is 
provided externally.
 If external track seed information is available -- then master module organizes 
search of candidates inside limited number of nearest signals in reference planes.

 In order to keep processing time reasonably small it's possible to split data from
 barrel tracker into few ``uncorrelated'' segments which have ``small'' 
 overlapping (because of bending of
 trajectories at low momenta). It will help essentially  because  
 number of combinatorial checks is proportional to the $N^{3}$ (where N is number
 of tracks or hits per layer).

\section*{Acknowledgments}
 The authors are grateful to W. Busza for support.
This work was partially supported by U.S. DoE grant
DE-FC02-94ER40818.

\end{document}